%% file: paper.tex
\documentclass[11pt]{article}

\usepackage{geometry}
\usepackage{graphicx}
\usepackage{color}
\usepackage{array}
\usepackage{amssymb,amsmath,amsfonts}
\usepackage{subfig}
\usepackage{tikz,pgfplots}
\usepackage{bm}
\usepackage{enumitem}
\usepackage[normalem]{ulem}
\usepackage{soul}
\usepackage{url}

\title{Functional Ratings in Sports}
\author{Bradley Lowery, Abigail Slater, and Kaison Thies \\ University of Sioux Falls
\thanks{This work was supported by the University of Sioux Falls Natural Science Research Fellowship Grant. A special thanks to Dr. Curt Buchholz and Dr. Carole Buchholz for supporting the grant.} 
}
\begin{document}
\maketitle

\begin{abstract}
\input{abstract}
\end{abstract}

\section{Introduction} 
\input{section-introduction} 
\section{Model}\label{sec.model}
\input{section-model}
\section{Data Set}
\input{section-data-set}

\newpage
\section{Comparing the Three Models}
\input{section-comparing-three-models}

\section{Functional Ratings}\label{sec.funRating}
\input{section-functional-ratings}

\section{Conclusion} 
\input{section-conclusion}

\newpage
\bibliographystyle{plain}
\bibliography{biblio}

\end{document}

%% file: abstract.tex
%abstract

In this paper, we present a new model for ranking sports teams.  Our model uses all scoring data from all games to produce a functional rating by the method of least squares.  The functional rating can be interpreted as a teams average point differential adjusted for strength of schedule.  Using two team's functional ratings we can predict the expected point differential at any time in the game.  We looked at three variations of our model accounting for home-court advantage in different ways.  We use the 2018-2019 NCAA Division 1 men's college basketball season to test the models and determined that home-court advantage is statistically important but does not differ between teams. 

%% file: section-introduction.tex
%Introduction 

Developing computer ranking systems for sports has been a popular problem for many years.  Computer rankings can be a valuable tool for ranking teams in leagues with many teams and only a limited number of games are played.  NCAA Division 1 men's college basketball is such a league.  In 2018-2019, there were 353 NCAA Division 1 teams and each team played about 30 games.  Ranking teams this way can be used for post-season selection while eliminating bias towards traditionally strong teams.   Rankings can also be used to generate win probabilities between any two teams regardless if they have played or not.   

Stefani~\cite{stefani.IEEE.1977} proposed using a linear model and the least squares solution to rank teams.  Many variations of the least squares method have been proposed~\cite{harville.JASA.1980,harville.AS.1994,massey1997,stefani.IEEE.1980} and our model is another variation using least squares.  Previous models used only final scores to generate rankings.  We propose using all scoring data to create a rating at any time in the game.  Similar rankings can be produced from this model.  Our model accounts for performance throughout the game rather than just the final outcome.  Additional information is obtained, such as the period of a game when a team is at their best or worst. 

%% file: section-model.tex
%Model

\subsection{Basic Model}

Suppose a set of $n$ teams play a total of $m$ games and we are interested in determining a ranking for 
the teams based not only on the final outcome of the game but also the point differential throughout the game.  

Let $T$ be the length of a game, $s_i(t)$ be team $i$'s score at time $t \in [0,T]$, and 
$\beta_i(t)$ be team $i$'s functional rating.    
If team $i$ plays team $j$, the model seeks to equate the difference of their functional ratings to the point 
differential ($d(t) = s_i(t) - s_j(t)$) in the game.  The model equation for a game is 
\begin{equation}\label{eq.model1}
\beta_i(t) - \beta_j(t) = d(t). 
\end{equation}

In the model, $\beta_1(t),\ldots,\beta_n(t)$ are unknown parameters that we will estimate using the 
method of least squares.  
Let $\bm{\beta}(t) = (\beta_1(t),\ldots,\beta_n(t))^\top$, $\bm{d}(t) = (d_1(t),\ldots,d_m(t))^\top$,
and $X$ be a $m\times n$ design matrix.  We use the notation that $x^{}_{ki}$ is the element in the $k$-th row and 
$i$-th column of $X$ and define $X$ such that 
\[x^{}_{ki} = \begin{cases}
1 , & \text{if $i$ is home team for game $k$,} \\
-1 ,& \text{if $i$ is away team for game $k$,} \\
0 , & \text{otherwise.}\\
\end{cases}\]
It does not matter which team is designated the home team if a game is played on a neutral site.  We will address home-court advantage in Section~\ref{sec.mod}.  
The linear system of equations can be expressed compactly as 
\[ X \bm{\beta}(t) = \bm{d}(t).\]

The model is a natural extension of traditional least squares rating systems~\cite{harville.JASA.1980,harville.AS.1994,massey1997,stefani.IEEE.1977,stefani.IEEE.1980} that use a linear model to relate team ratings to the outcome of the games.  In those models, the dependent variable is a scalar quantity, such as margin of victory or a score that is dependent on the margin of victory.  In our model, the dependent variable is a function of time, so we obtain a rating for each team as a function of time.  We will discus how to use the functional rating to determine a ranking in Section~\ref{sec.func2scalar}, but the functional rating will also be of interest to us.   

We will demonstrate that most of the modifications and analysis for the scalar models can be replicated for the functional models.  We will also show how to use the functional ratings to gain additional insight on
the teams.  

\subsection{Modifications}\label{sec.mod}
There are two different modifications to account for home-court advantage.  Our first modification
assumes the home-court advantage is the same for all teams.  The model equation for a game is 
\begin{equation}\label{eq.model2}
	d(t) = 
	\begin{cases} 
		\beta_i(t) - \beta_j(t) + \alpha(t), & \text{if team $i$ is the home team}  \\
		\beta_i(t) - \beta_j(t), & \text{if the game is played on a neutral site.}  \\
	\end{cases}
\end{equation}
The difference between two team's functional ratings is the expected point differential on a neutral site and 
$\alpha(t)$ is the additional points for the home team if played on their court. 
The design matrix for this model is obtained by adding a column to the design matrix of the basic model. Let $X$ be the $m\times (n+1)$ design matrix.  The first $n$ columns are the same as in the basic model and the $(n+1)$-th column is defined by 
\[x^{}_{k(n+1)} = \begin{cases}
0 , & \text{if game $k$ is played on a neutral site,} \\
1 ,& \text{if game $k$ is played on team $i$'s home-court,} \\
\end{cases}\]
If we define $\bm{\beta}(t) = (\beta_1(t),\ldots,\beta_n(t),\alpha(t))^\top$, the matrix equation has the same form as the 
basic model 
\[ X \bm{\beta}(t) = \bm{d}(t).\]

The second modification has a specific home-court advantage for each team, 
\begin{equation}\label{eq.model3}
	\beta_i(t) - \beta_j(t) + \alpha^{}_i(t)= d(t).
\end{equation}
The design matrix for this model is again obtained by adding columns to the design matrix of 
the basic model. Let $X$ be the $m\times (2n)$ design matrix.  The first $n$ columns are the same as in the basic model and the $(n+i)$-th column is defined by 
\[x^{}_{k(n+i)} = \begin{cases}
1 ,& \text{if game $k$ is played on team $i$'s home-court,} \\
0 , & \text{otherwise.} \\
\end{cases}\]
The case when $x_{k(n+i)} = 0$ includes when game $k$ is a neutral site game and when
team $i$ is not playing the game.  
If we define $\bm{\beta}(t) = (\beta_1(t),\ldots,\beta_n(t),\alpha_1(t),\ldots,\alpha_n(t))^\top$, then matrix equation again has the same form as the 
basic model 
\[ X \bm{\beta}(t) = \bm{d}(t).\]

These modifications have been used in the past for ranking systems~\cite{harville.AS.1994,stefani.IEEE.1980}.  
Again, the main difference in our model is the dependent variables and parameters are functions rather than scalars.    

\subsection{Overtime Games}

All of the models assume each game has the same length, which poses a problem for overtime games.  We decided to remove overtime data and therefore, overtime games end in a tie. Another option could have been to extend all games to the longest game in the data set, however, this may give undesirable weight to the final score since games that ended in regulation would record the final score for all possible overtime periods.  Further study is need to determine how to incorporate overtime data in the model.

\subsection{Constraints} 

	The design matrix in the basic model is not full rank because the null space has dimension of at least 1, which is readily seen because $X\bm{v} = 0$ when $\bm{v} = (1,1,1,...,1)^\top$.  The design matrix for the other models are also not full rank by a similar argument.  Therefore, there is no unique least squares solution, thus we are free to add a constraint. We use the constraint that the average team rating is the zero function, that is 
			\begin{equation}
			\sum_{i = 1}^{n}\beta_i(t) = 0, \; \forall \; t.
			\end{equation}
			
	In general, we can choose a constraint that would set any team's rating to a constant of our choice. This will set teams' ratings so that they are relative to the chosen constraint, and is just a shift of the ratings, with no impact on the rankings. A logical choice could be to set the worst team to be the zero function so that all other teams' ratings are positive.
	Another option is to set the constraint to be the average score of all of the games for any time, which will shift our ratings to be interpreted as the average score for a team at any time.

	The dimension of the null space of $X$ is exactly 1 if enough games have been played so all of the teams are connected.  If not enough games are played then a rating comparing all teams is not possible and one can only compare teams within a connected subset.  We will always assume enough games are played so a rating comparing all teams is valid.  

\subsection{Least Squares Solution} 

Define the {\it residual} vector-valued function to be  $\bm{r}(t) = X\bm{\beta}(t) - \bm{d}(t)$.   The least squares solution seeks to minimize the $L^2$-norm of the residual, 
\[ || \bm{r}(t) || = \int_{0}^{T} \sum_{i=1}^{n} r_i^2(t) dt 
			= \int_{0}^{T} \bm{r}(t)^\top \bm{r}(t) dt
			= \int_{0}^{T} (X\bm{\beta}(t) - \bm{d}(t))^\top (X\bm{\beta}(t) - \bm{d}(t)) dt. 
\] 

Ramsay and Silverman~\cite[Chapter~13]{ramsay.springer.2005} discuss various solution methods.   
We use pointwise minimization on the raw data which is discretized at every second in the game.  The solution for this approach is obtained by finding the traditional least squares solution at each second.  By using the raw data there is no loss of information prior to minimizing.   In Section~\ref{sec.funRating}, we will smooth the functional ratings for easier interpretation of the results.

%% file: section-data-set.tex
%Data Set Section

In the following sections, the linear models are fit using the 2018-2019 NCAA Division 1 men's college basketball season.  Including the postseason, there are a total of 5603 games.
For each game, we have the high level information which includes - the date that game occurred, the home and away team, the final score, and whether or not it was played at a neutral site. We also have a scoring summary for each game that consists of each time there is a score change and each teams score at that specific time in the game.  From the scoring summary, we are able to interpolate to obtain each team's score at every second in the game.

We were unable to find a single source that provided all of the scoring summaries.
We used SportsReference.com~\cite{sports.reference} to get most of the scoring summaries and then ESPN.com~\cite{espn} to get as many of the other scoring summaries as possible. Some scoring summaries were not available on either Sports Reference or ESPN.  In those cases, we obtained the scoring summary from the school's website of one of the teams that played in the game. 

We verified we had a complete data set by comparing the high level game information to another data set. Kenneth Massey, an American Sports Statistician, has a complete set of high level information for each game on MasseyRatings.com~\cite{masseyratings}. Massey's data includes who was playing, the final score, the date of the game, and whether or not it was played at a neutral site.  After verification, our data set contains the same high level game information as Massey's data set.
 
\subsection{Errors}   
The data set contains three types of errors. The first error is for only one game, Jackson State at Alabama A\&M on January 5, 2019. This game did not have a public scoring summary on any website. Therefore, this scoring summary has only has four data points obtained from the box score: the beginning of the game, the halftime score, the end of regulation score, and the end of overtime score. The second error comes from the final entry in the scoring summary not matching the correct final game score. In these cases, the scoring summary was missing a scoring play but we were unable to find the exact time in the game where the scoring play was missed. To ensure the final score was correct, we added the missing score as the last data point in our scoring summary.  

\newpage

The final error is when multiple scoring plays occur at the same time.  For example, Table~\ref{tab.scoringsummary} shows the first ten scoring plays for North Alabama at Samford on November 6, 2018.  The play-by-play indicates that North Alabama makes two free throws at 40~s into the game, play continues with multiple missed shots, rebounds, and other scoring plays while the time remains at 40~s.  Finally the time is updated when North Alabama makes a layup at 253~s into the game.  The team's scoring functions show North Alabama having 7 points from 40~s to 252~s and Samford having 12 points over that same time period.  The result in the model is that Samford is credited with a 5 point advantage over the entire time interval, when they are actually only winning by 5 points on a shorter time interval.   Since we are unable to account for when the actual scoring plays happened we did not modify these situations.

\begin{table}
\parbox{.45\textwidth}{
\begin{center}
\begin{tabular}{c|c|c}
Time (s) & North Alabama & Samford \\\hline
30    &     2    &     0\\
40    &     3    &     0\\
40    &     4    &     0\\
40    &     4    &     2\\
40    &     4    &     5\\
40    &     4    &     7\\
40    &     7    &     7\\
40    &     7    &    10\\
40    &     7    &    12\\
253   &     9    &    12
\end{tabular}
\end{center}
\caption{First ten scoring plays for North Alabama at Samford on November 6, 2018.}
\label{tab.scoringsummary} 
}
\hspace{.5in}
\parbox{.45\textwidth}{
\begin{center}
\begin{tabular}{c|c|c}
Time (s) & LSU & Florida \\\hline
2159   &     63    &    60\\
2175   &     63    &    63\\
2175   &     63    &    64\\
2175   &     63    &    65\\
2175   &     63    &    66
\end{tabular}
\end{center}
\caption{A few scoring plays for LSU vs. Florida on March 15, 2019.}
\label{tab.scoringsummary2} 
}
\end{table}

\paragraph{Note.} Not all scoring plays that happen at the same time occur in error.  This will be the situation
for free throws since the clock is not running during the free throw attempt.  Large score changes at a given time are possible if a basket is made and a foul is called on the play. For example, Table~\ref{tab.scoringsummary2} shows a few scoring plays for LSU vs. Florida on March 15, 2019. 
At 2175~s into the game Florida made a three point shot and a foul on LSU occurred during the play.  
There was also a technical foul called on LSU resulting in a total of four free throw attempts for Florida.  Florida made three of the four free throws resulting a six point play.

%% file: section-comparing-three-models.tex
%Comparing-Three-Models

% do we need to add the equation for the F test
%k is the full model (number of teams +1)
%m is number of teams (so df is 1)
%n is num of games

\setlist[itemize]{noitemsep, topsep=5pt}
\setlist[enumerate]{noitemsep, topsep=5pt}
Recall from Section~\ref{sec.model} our three models are
\begin{itemize}
	\item[] Model 1 -- no home-court advantage~\eqref{eq.model1};
	\item[] Model 2 -- constant home-court advantage~\eqref{eq.model2};
	\item[] Model 3 -- individual home-court advantage~\eqref{eq.model3}.
\end{itemize}

To determine which model is most appropriate for our data, we use Analysis of Variance (ANOVA) similar to Harville and Smith~\cite{harville.AS.1994}.  The test compares the sum of square error of the {\it full} model ($\text{SSE}_f$) to the sum of square error of the {\it reduced} model ($\text{SSE}_r$).  A reduced model is the same as the full model with a constraint added on some parameters.  If $\text{SSE}_r$ is much larger than $\text{SSE}_f$ than the full model is needed, otherwise the reduced model is appropriate.  Details of the test are outlined in~\cite[Section 12.5]{milton.probandstat.2003}.  We conduct the same test at every second in the game resulting in a function of $P$-values.

\begin{figure}
	\parbox{.45\textwidth}{
	\begin{center} \includegraphics[width=3in]{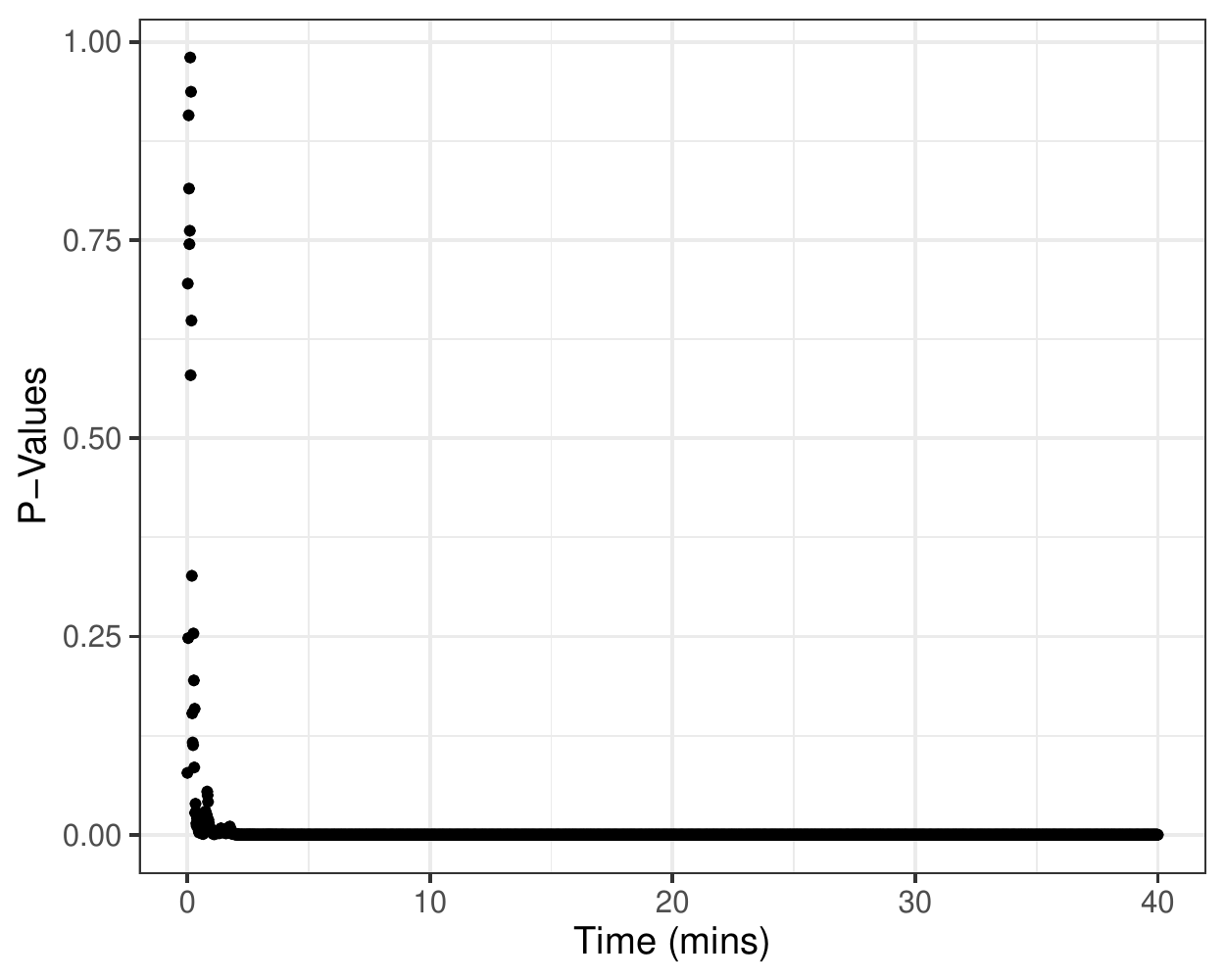} \end{center}
	\caption{$P$-values for Model 1 vs. Model 2.}
	\label{test1}}
	\hspace{.4in}
	\parbox{.45\textwidth}{
	\begin{center} \includegraphics[width=3in]{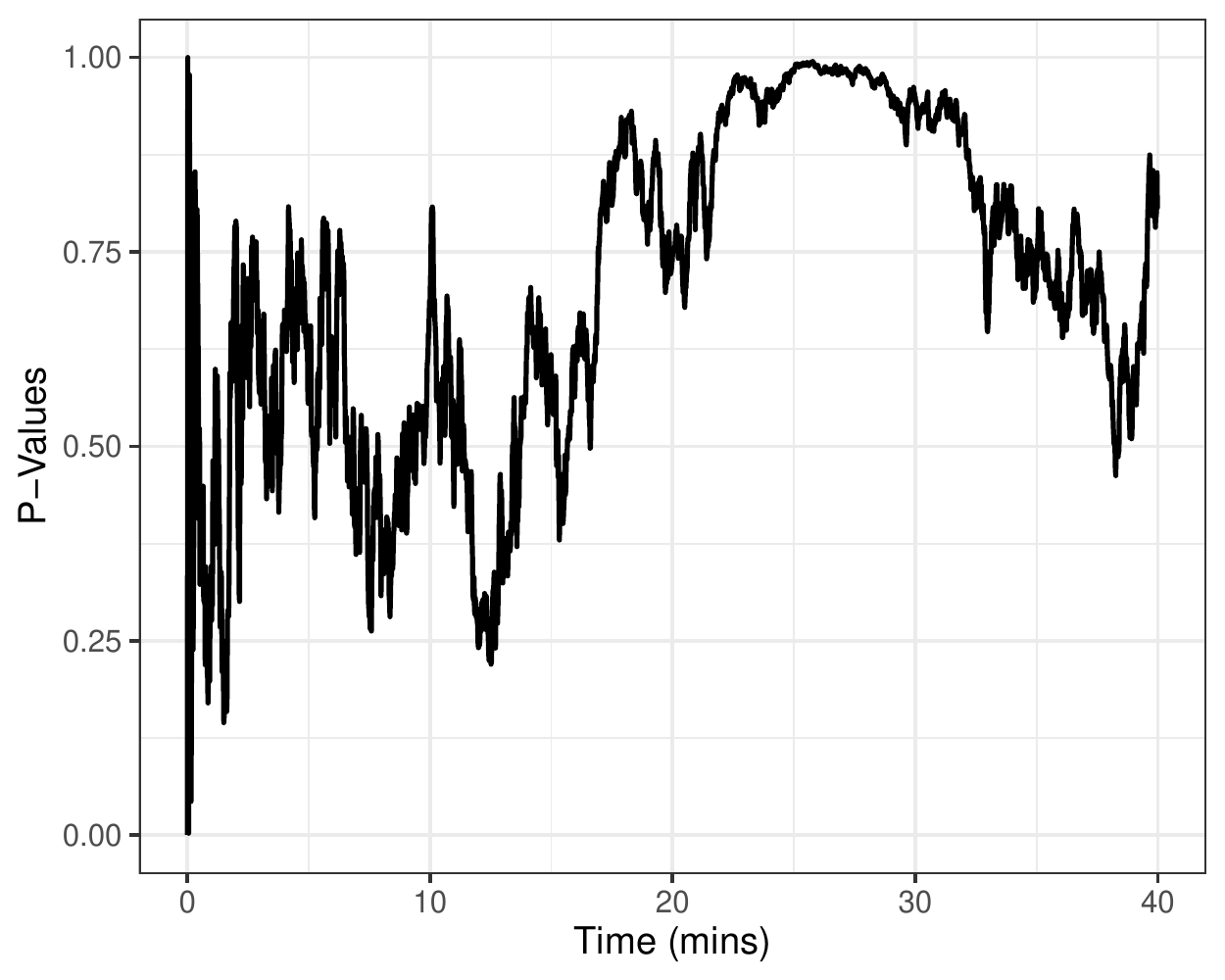} \end{center}
	\caption{$P$-values for Model 2 vs. Model 3.}
	\label{test2}}
\end{figure}

First let us compare Model 1 and 2. Model 1 is a reduced version of Model 2, which is readily seen by setting $\alpha(t) = 0$ in~\eqref{eq.model2} and obtaining~\eqref{eq.model1}. The hypotheses for this test at a given time $t$ are as follows:
\begin{itemize}
	\item[] $H_0$: $\alpha(t) = 0$ (no home-court advantage model is appropriate);
	\item[] $H_1$: $\alpha(t) \neq 0$ (constant home-court advantage model is needed).
\end{itemize}
Figure~\ref{test1} shows the $P$-values for the test at each second in the game. Since the $P$-values are consistently less than .1 other than the first minute of the game, we reject the null hypothesis and conclude that the constant home-court advantage model is needed.

Next we wish to test if individual home-court advantage is appropriate.  Model 2 is the reduced model of Model 3 if we impose the constraint that all home-court advantages are equal.  The hypotheses for this test at a given time $t$ are as follows: 
\begin{itemize}
	\item[] $H_0$: $\alpha_1(t) = \alpha_2(t) = \cdots = \alpha_n(t)$ (constant home-court model is appropriate)
	\item[] $H_1$: at least one $\alpha_i(t)$ differs from the others (individual home-court model is needed)
\end{itemize}
Figure~\ref{test2} shows the $P$-values for the test similar to Figure~\ref{test1}. Since the $P$-values are consistently greater than .1 other than the first 20 seconds, we fail to reject the null hypothesis and conclude that the constant home-court advantage model is appropriate. The conclusion of both tests is that Model~2 is appropriate, thus we use Model 2 for the rest of the paper.

\begin{figure}
	\begin{center} \includegraphics[width=3in]{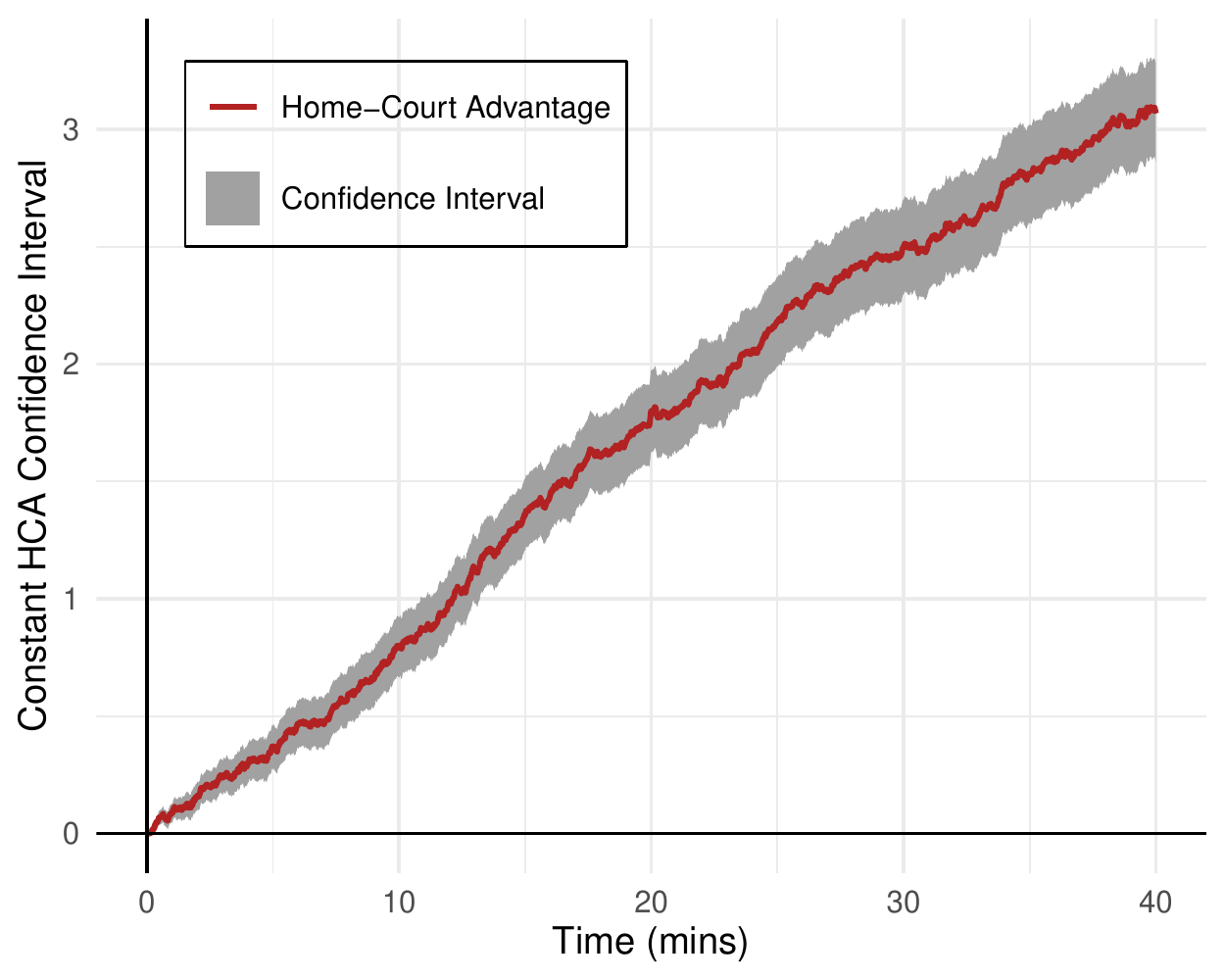} \end{center}
	\caption{Constant home-court advantage function ($\alpha(t)$) with an 80\% confidence interval.}
	\label{conf-Int}
\end{figure}

The confidence interval for constant home-court advantage can be helpful to determine if it is significant enough to be of practical importance.  As seen in Figure~\ref{conf-Int}, the 80\% confidence interval for the constant home-court advantage function is reasonable.  The home team can expect to have an end of the game advantage of about three points, which would be significant for two closely rated teams, but not so drastic that home-court advantage would determine the outcome of a game for two unevenly matched teams. 

%Following is removed.  Irrelevant unless showing individual home-court.
%Note that even though we may expect a team to have a positive home-court advantage, this is not a requirement.  An individual team could have a negative home-court advantage, which simply means the team is performing worse at home than on the road. 

%% file: section-functional-ratings.tex
%Functional Ratings

\begin{figure}
	\parbox{.45\textwidth}{
		\begin{center} \includegraphics[width = 3in]{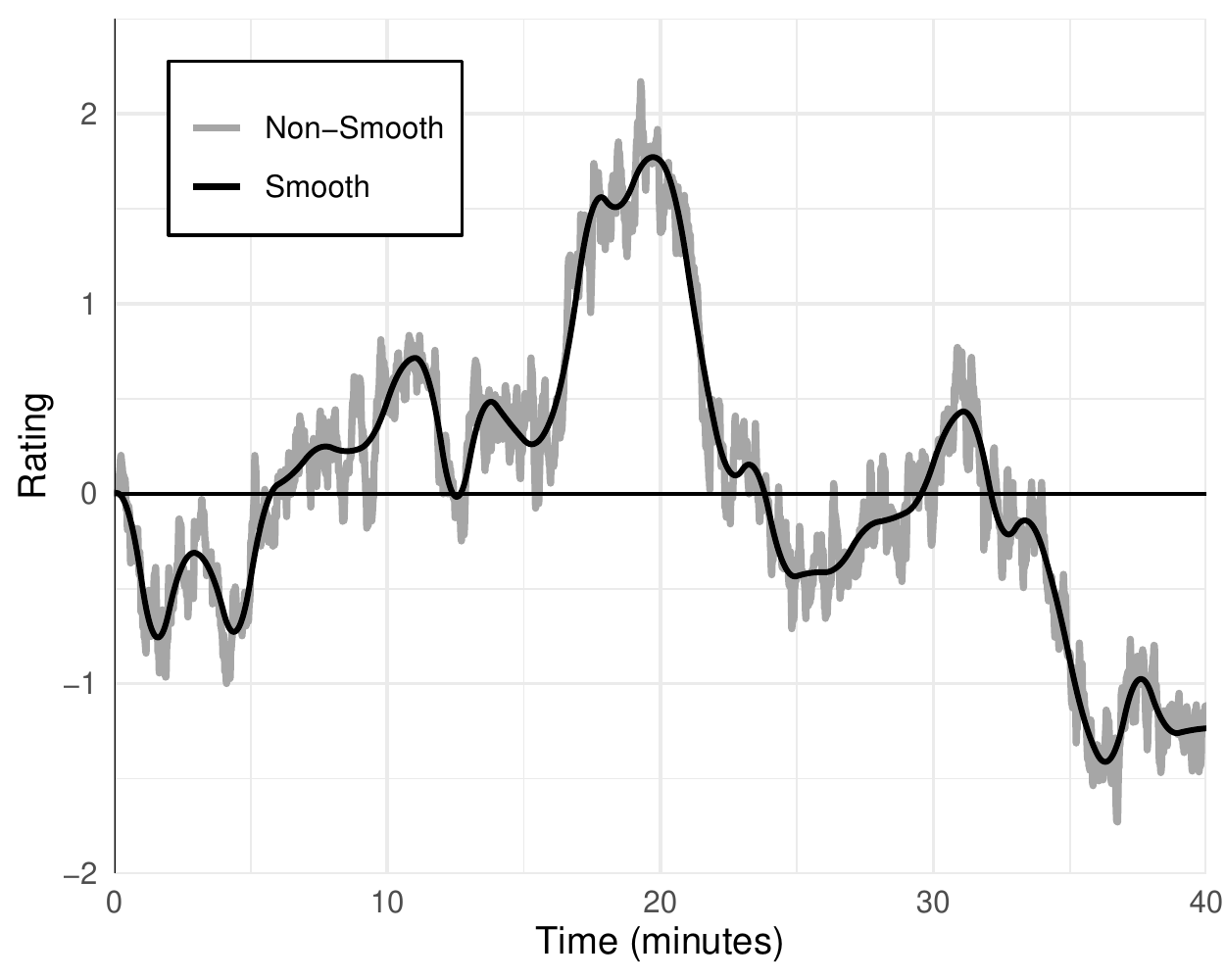} \end{center}
		\caption{Non-smooth and smooth functional rating for Louisiana.}
		\label{SmoothVnonSmooth}
	}
	\hspace{.4in}
	\parbox{.45\textwidth}{
		\begin{center} \includegraphics[width = 3in]{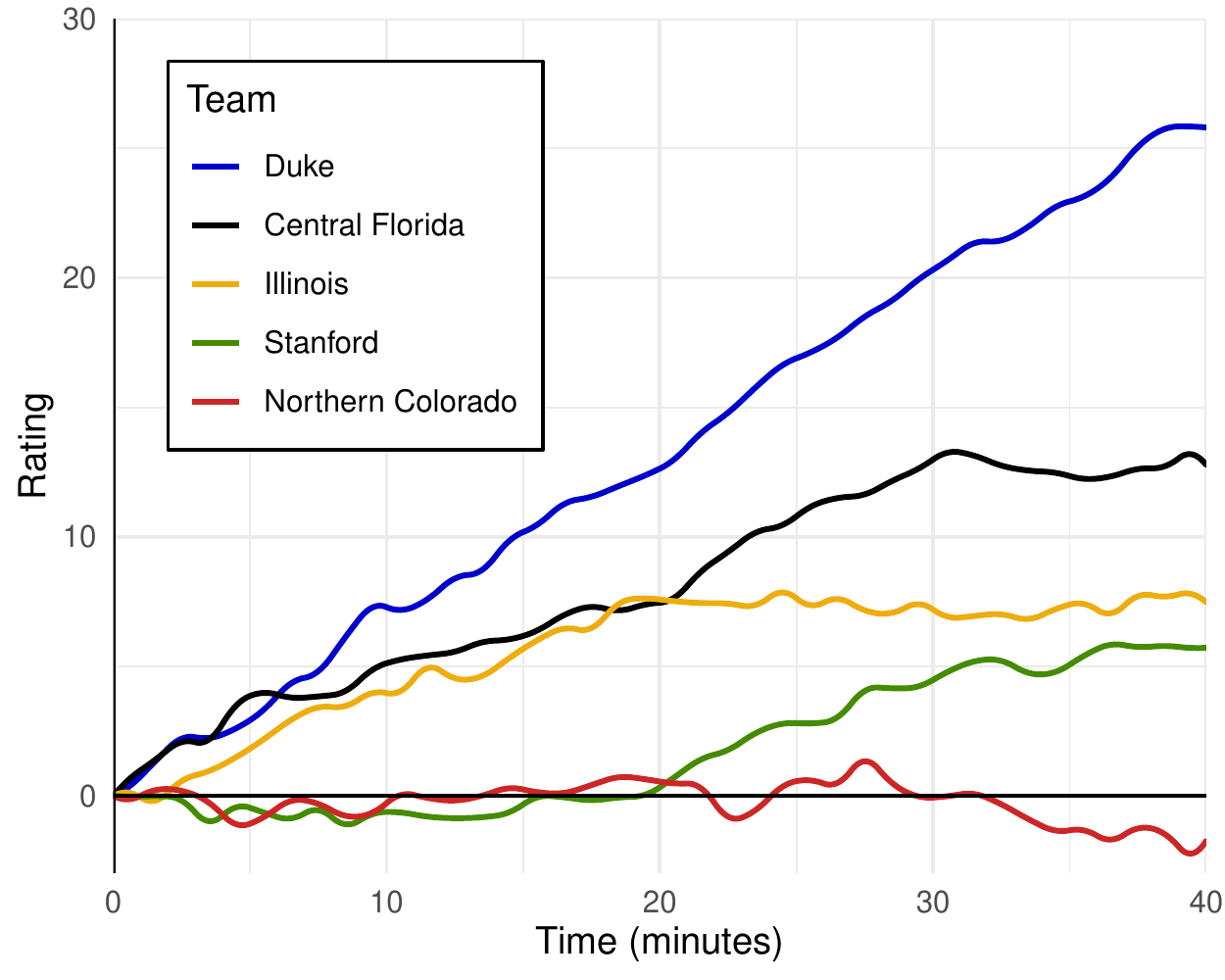} \end{center}
		\caption{Functional rating for five teams.}
		\label{fig.5teams}
	}
\end{figure}

The graph of the functional ratings are noisy, but a general trend is definitely discernible. Using smoothing techniques proposed by Ramsay and Silverman~\cite{ramsay.springer.2005}, we are able to eliminate the noise without losing too much information about the functional ratings. 
Via trial and error, we found that an order four B-spline with knots every minute creates a nicely smoothed graph with little loss of information.  Figure~\ref{SmoothVnonSmooth} shows the smooth and non-smooth functional rating for Louisiana. Comparing the smooth and non-smooth graphs, we see that there is little difference and small variations are still visible.

Figure~\ref{fig.5teams} shows five different team's functional rating. Duke is an example of a team whose rating is consistent throughout the game. Many high rated teams have this type of functional rating, but this type is not limited to only high rated teams. Middling teams can have a similar shape, but the general slope would be smaller than that of a team like Duke. Central Florida and Illinois show functional ratings often found in a second tier of teams. These ratings seem to level off during the last few minutes of a game.  Illinois is an extreme case as they level off for the entire second half.  In contrast, Stanford shows a team that is average for the first half and improves greatly during the second half.  Finally, Northern Colorado is a team that is fairly average throughout the entire game.

\subsection{Functional to Scalar Ratings}\label{sec.func2scalar} 
How can we create a ranking system based on our functional data? A logical choice is to find the (weighted) average functional rating for each team: 
\begin{equation}\label{eq.scalar}
	\frac{\int_{0}^{T} w(t)\beta_i(t)dt}{\int_{0}^{T}w(t)dt}
\end{equation}
Using this scalar value, we can put the teams in rank order. If $w(t)=1$ for all $t$, the scalar rating is simply the average rating over the entire game.  Of course, there are many different weights we could use, depending on what properties one values. For example, we could make the argument that the closer to the end of the game two teams are, the more impact the score differential should have on their rating (i.e. We want to favor teams who are better at the end of games).  A simple choice would be to use an increasing linear function, but we can expand this idea to any non-decreasing weight function. 

Table~\ref{Rankings} shows the ranking results for a few selected teams.  The table includes the scalar rating using $w(t)=1$ for all $t$ in~\eqref{eq.scalar}, the corresponding rank, and also the rank if only the end of the game rating was used to rank the teams. Comparing the two rankings, the top ten teams are the same, but in a different order.  More movement occurs for middling teams. Central Florida and Illinois both see a 20 position decrease when only looking at the end of the game.  This corresponds to their functional rating leveling off towards the end of the game as shown in Figure~\ref{fig.5teams}.  On the other hand, Stanford's ranking increases by 40 positions when only considering the end of the game.  Again, this corresponds to their functional rating improving greatly towards the end of the game. 

\begin{table}
	\parbox{.5\textwidth}{
		\begin{center}
			\begin{tabular}{r|c|c|c}
				Team            & \shortstack{Scalar \\ Rating} & Rank & \shortstack{End of \\ Game \\ Ranking}  \\\hline
				Gonzaga         &     14.98    &     1 &1\\
				Duke            &     13.31    &     2 &2\\
				Virginia        &     12.99    &     3 &3\\
				North Carolina  &     12.77    &     4 &5\\
				Michigan        &     12.45    &     5 &7\\
				Michigan State  &     11.87    &     6 &4\\
				Tennessee       &     11.71    &     7 &9\\
				Purdue          &     11.63    &    8  & 10\\
				Texas Tech      &     11.30    &    9  &6\\
				Kentucky        &     10.64    &    10 &8\\
				Central Florida &      8.16    & 22    &40\\
				Illinois        &      5.58    & 52    &72\\
				Stanford        &      1.78    & 132   &92\\
				Northern Colorado &   -0.24    & 167   &191\\
				Chicago State   &    -12.89    & 353   &352
			\end{tabular}
			\caption{Select team's scalar rating using~\eqref{eq.scalar} with $w(t)=1$, the corresponding rank, and the rank based on their end of the game rating.}
			\label{Rankings} 
		\end{center}
	}
%\end{table}
\hspace{.4in}
%\begin{table}
	\parbox{.4\textwidth}{
		\begin{center}
			\begin{tabular}{r|c|c}
				Team            & \shortstack{Scalar \\ SOS}  & Rank \\\hline
				Kansas            &  6.10   &    1  \\
				Michigan State    &  6.00   &    2  \\
				Purdue            &  5.91   &    3  \\
				Oklahoma          &  5.87   &    4  \\
				Duke              &  5.80   &    5  \\
				Penn State        &  5.72   &    6  \\
				North Carolina    &  5.63   &    7  \\
				Minnesota         &  5.60   &    8  \\
				Florida           &  5.58   &    9  \\
				Nebraska          &  5.48   &   10  \\
				Illinois          &  5.34   &   13  \\
				Central Florida   &  2.75   &   62  \\
				Stanford          &  2.18   &   75  \\
				Northern Colorado & -2.71   &  316  \\
				Morgan State      & -5.01   &  353
			\end{tabular}
			\caption{Scalar strength of schedule computed using~\eqref{eq.scalar} with $w(t)=1$ for some select teams.} 
			\label{tbl.sos} 
		\end{center}
	}
\end{table}

\newpage

\subsection{Interpreting Functional Ratings} 

A nice interpretation of the functional ratings is obtained by examining the normal equation, 
\[X^T X \bm{\beta}(t) = X^T \bm{d}(t).\]
Solving equation $i$ for team $i$'s functional rating shows that it can be slit into two parts:
\begin{equation}
	\beta_i(t) = \bar{d}_i(t) + sos_i(t),
\end{equation}
where $\bar{d}_i(t)$ is the team's average point differential and $sos_i(t)$ is the team's strength of schedule. 

Team $i$'s {\it average point differential} is defined to be, 
\begin{equation}
\bar{d}_i(t) = \frac{1}{m_i}\sum_{k \in G_i}x_{ki}d_k(t),
\end{equation}
where $G_i$ is the set of games that team $i$ played and $m_i$ is the number of games team $i$ played. Recall 
that $x_{ki} = 1$ if team $i$ is the home team and $x_{ki} = -1$ if team $i$ is the away team.  The result is $x_{ki}d_k(t)$ will always treat the point differential for game $k$ relevant to team $i$ (positive point differential means team $i$ is winning).

Team $i$'s {\it strength of schedule} is defined to be, 
\begin{equation}
\text{sos}_i(t) = \frac{1}{m_i}\sum_{j \in T_i}\beta_j(t) - \frac{h_i}{m_i}\alpha(t) + \frac{a_i}{m_i}\alpha(t)
\end{equation}
where $T_i$ is the set of teams that played team $i$ (if a team is played multiple times they would be listed for each game played), $h_i$ is the number of home games, and $a_i$ is the number of away games. The first term of $\text{sos}_i(t)$ is the average rating of team $i$'s opponents, the second term is decreasing the strength of schedule for the fraction of games played at home (indicating an easier schedule), and the last term is increasing the strength of schedule for the fraction of games played on the road (indicating a harder schedule).  The result is if a team plays more home games than away their strength of schedule will be less than their average opponents rating (or vice versa).  Table~\ref{tbl.sos} shows the scalar strength of schedule for some select teams.  The scalar strength of schedule is computed using~\eqref{eq.scalar} with $w(t)=1$ and $\text{sos}_i(t)$ rather than $\beta_i(t)$.   As expected, the top 10 strength of schedules are all from major conferences (Big 12, Big 10, and ACC).

\begin{figure}
	\parbox{.45\textwidth}{
		\begin{center} \includegraphics[width = 3in]{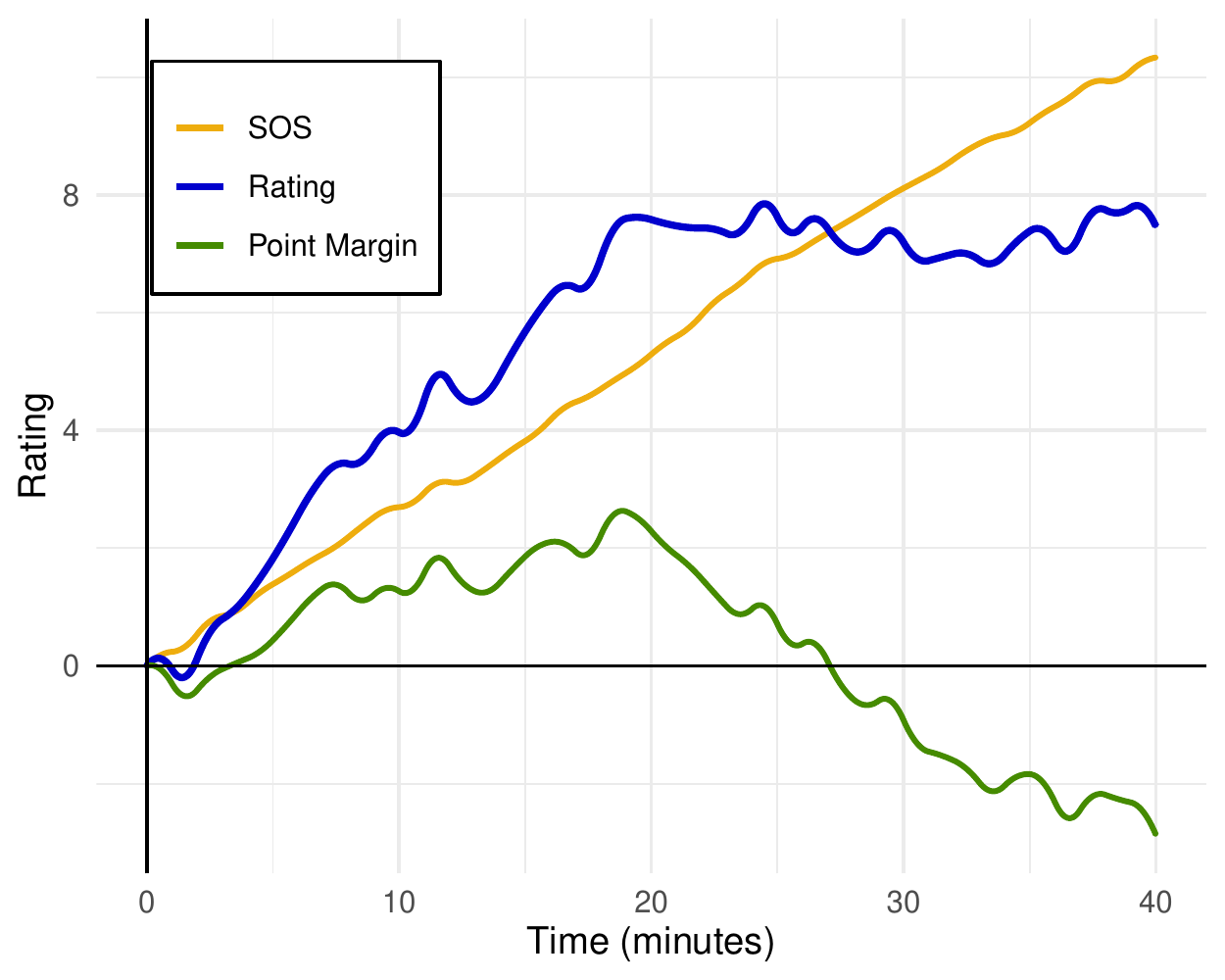} \end{center}
		\caption{Illinois functional rating, strength of schedule, and average point differential.}
		\label{fig.illinois}
	}
	\hspace{.4in}
	\parbox{.45\textwidth}{
		\begin{center} \includegraphics[width = 3in]{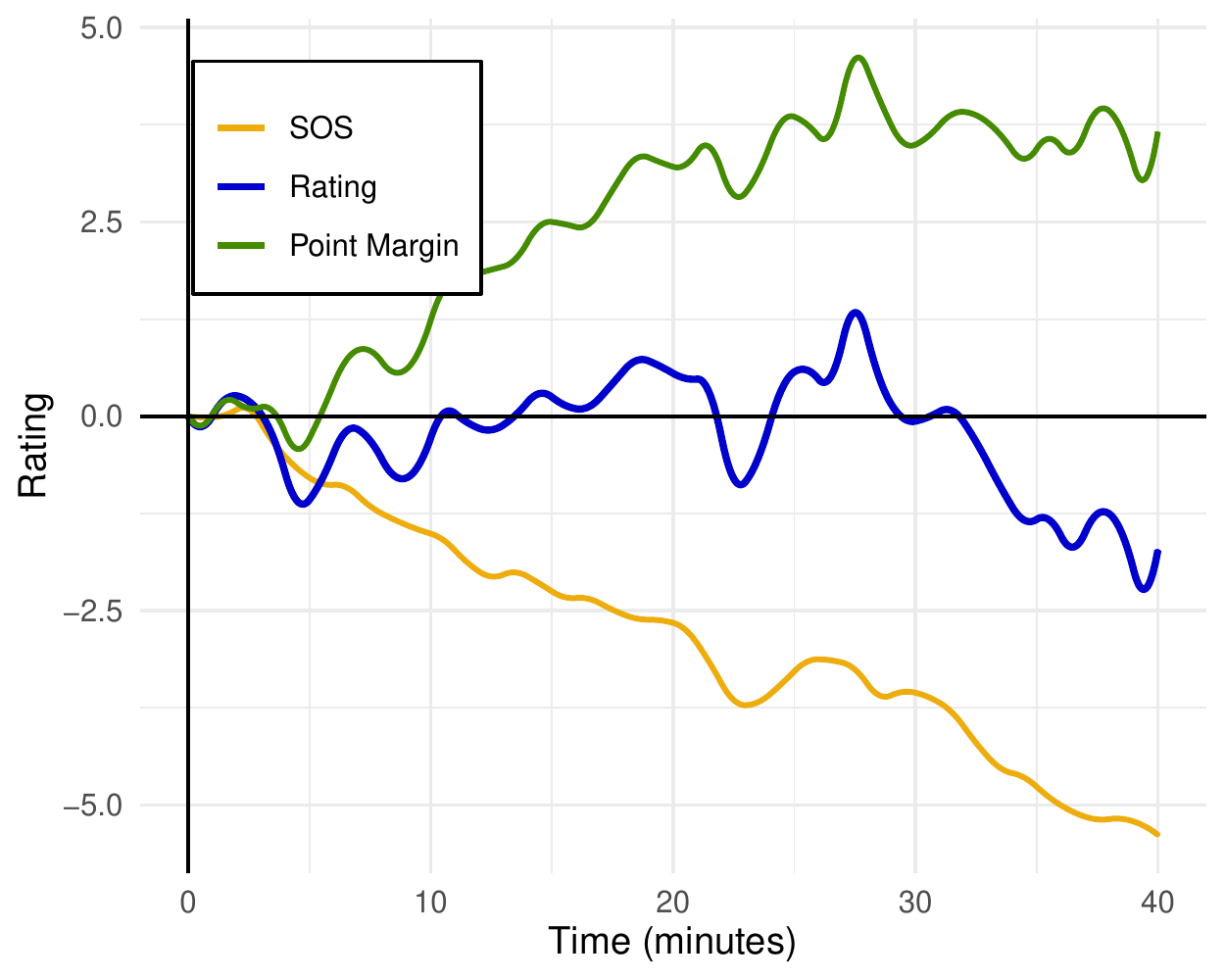} \end{center}
		\caption{Northern Colorado functional rating, strength of schedule, and average point differential.}
		\label{fig.unc}
	}
\end{figure}

Figure~\ref{fig.illinois} shows the strength of schedule, average point differential, and functional rating for Illinois.  Illinois has a unique functional rating since it is roughly constant after halftime.  Breaking down the rating into strength of schedule and average point differential shows why this happens.   On average, Illinois is leading games for the first half and then trailing for the second half.  However, the decrease in point differential during the second half is countered by the increasing strength of schedule resulting in a roughly constant functional rating.  The inflection point in the functional rating occurs at the same time as the inflection point in the average point differential because the strength of schedule is increasing fairly consistently.  This is a trend for most strength of schedules and the result is inflection points in the functional rating correspond to inflection points in the average point differential.  

Figure~\ref{fig.unc} shows the same graph for Northern Colorado.  Northern Colorado is an example of a winning team (positive point differential and 19-11 overall record) that plays a weak schedule (negative SOS). The result is their rating is roughly average (167 out of 353).  This example clearly demonstrates that strength of schedule is a factor in determining a team's rating. 

\begin{figure}
	\begin{center} \includegraphics[width = 3in]{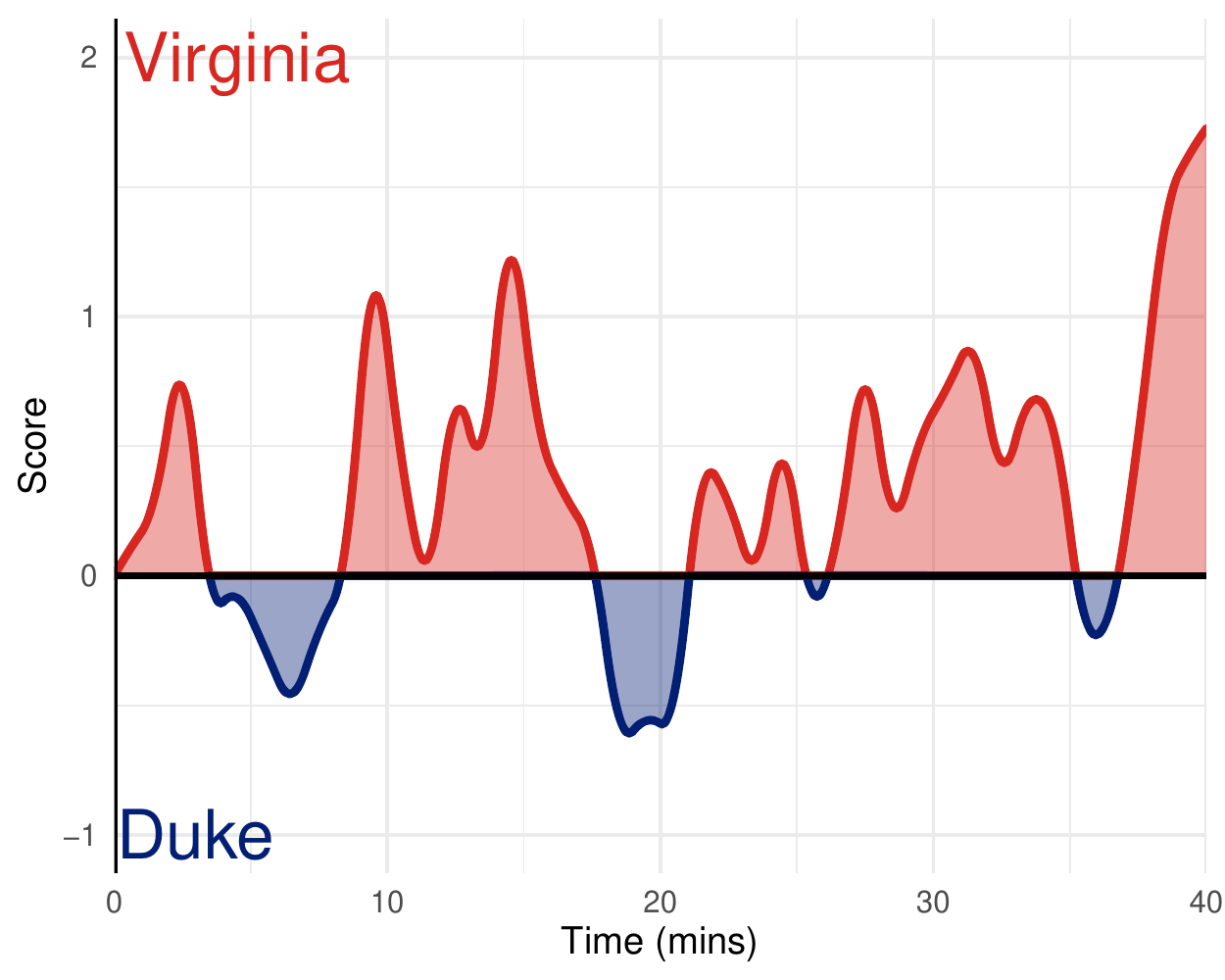} \end{center}
	\caption{Predicted game between Duke and Virginia of which is filled to be their team color if winning. }
	\label{DukevsVirginia}
\end{figure}

\newpage

\subsection{Predictions}

By simply subtracting each team's functional rating at every time, we get a prediction at a neutral site between these two teams. Figure~\ref{DukevsVirginia} shows an example of a prediction for Virginia versus Duke (two highly rated teams). We notice that the predictions are probably not the most accurate representation of an actual game; as an actual game would typically have larger runs for each team.  Our model predictions represent the average point differential if many games were played between the two teams. Further study is needed to determine if the functional ratings can be used to simulate an actual game.

%% file: section-conclusion.tex
%Conclusion

Functional linear regression can be used to determine a functional rating for teams.  Similar analysis used for scalar linear regression models is possible while gaining additional information.  A team's functional rating can be interpreted as a combination of the team's average point differential and their strength of schedule.  With this interpretation the functional rating can provide insight to a team's playing style, such as being a strong first or second half team and what level of competition they play.  Functional ratings can be used to predict the flow of a game, but the predictions are not accurate to a realistic game since the prediction is unable to capture large runs for either team as typically seen in games.  

This paper provides the initial framework for working with functional ratings.  Future work may include applying the functional rating to other sports where scoring more discrete, such as football or soccer; using the  ratings to simulate realistic games; and computing win probabilities.